# Inefficiency in selecting products for submission to national research assessment exercises[1]

*Giovanni Abramo[a,c\*], Ciriaco Andrea D'Angelo [a,b], Flavia Di Costa[a,c]*

[a] Laboratory for Studies of Research and Technology Transfer
Institute for System Analysis and Computer Science (IASI-CNR)
National Research Council of Italy

[b] Department of Engineering and Management
University of Rome "Tor Vergata"

[c] Research Value S.r.l.

**Abstract**

One of the critical issues in national research assessment exercises concerns the choice of whether to evaluate the entire scientific portfolio of the institutions or a subset composed of the best products. Under the second option, the capacities of the institutions to select the appropriate researchers and their best products (the UK case) or simply the best products of every researcher (the Italian case) becomes critical, both for purposes of correct assessment of the real quality of research in the institutions evaluated, and for the selective funding that follows. In this work, through case studies of three Italian universities, we analyze the efficiency of the product selection that is intended to maximize the universities' scores in the current national research assessment exercise, the results of which will be the basis for assigning an important share of public financing over the coming years.

---

# 1. Introduction

Higher education institutions (HEIs) are an important pillar of national research and innovation systems, thus the policy agendas of many countries now place high priority on strengthening such institutions. One expression of this policy is the increasing diffusion of national research assessment exercises. Beginning from the original British RAE, which has inspired many other countries, national assessment exercises are becoming regular events, particularly in nations wishing to introduce new principles of governance and more innovative management of the research sphere, such as through performance-based research funding systems (PBRFs) – which in turn depend on the results of the national assessments. The intended purpose of PBRFs is to influence improvement in underperforming institutions (Herbst, 2007, p. 90) and to stimulate continuous improvement of productivity in the whole research system (Abramo et al., 2009).

An international comparative analysis on the adoption of PBRFs, by Hicks (2012), indicates that following the example of the RAE, at least 14 other countries (of which 11 in the EU, plus China, Australia, New Zealand) now base some portion of public financing for research institutions on national assessment exercises. The magnitude of the funds assigned on the basis of the evaluations varies from nation to nation, as does the methodology for assigning them. Yet the adoption of PBRFs does not seem immune from serious concerns and risks, as shown in a study by Geuna and Martin (2003) analyzing the experience of twelve countries in Europe and the Asia-Pacific region. Their analysis indicates that while the initial benefits of PBRFs might outweigh the costs, they produce diminishing returns over time. In any case, national evaluations continue, taking on particular importance where the results are used to support important policy decisions or apply incentives that influence the behavior of scientists (Council of Canadian Academies, 2012). Indeed, a typical assumption for any national evaluation process is that it is intended to produce information and data to "enhance management practices and achieve goals, provide accountability for the stated goals, improve performance, allocate resources, and lead to informed policy decisions" (Jordan and Malone, 2001). National assessments serve not only to allocate resources in an efficient manner and stimulate improvement in performance, but also to reduce information asymmetry between supply and demand for new knowledge, and further to demonstrate that investment in research is effective and delivers public benefits. Given that assessment exercises are intended to pursue such important objectives, it is clear that they must be executed with maximum methodological rigor, applying the best technologies available.

Planning of national research assessment exercises has always involved a choice between two distinct methodologies: peer review and bibliometry. Until recently, the most commonly adopted method was peer review, where products submitted by institutions are evaluated by panels of appointed experts. But new developments in bibliometric indicators have led many governments to introduce the use of these metrics, either in integration with peer review (known as "informed peer review") or in complete substitution. The use of bibliometric measures is still limited to the hard sciences, where publications in international journals and conference proceedings are the most accepted form for the codification of new knowledge, and where the publications therefore represent a trustworthy proxy of overall research outputs.

Both approaches offer their own strengths and weaknesses, which have been amply discussed in the literature (Abramo and D'Angelo, 2010; Moxham and Anderson 1992; Horrobin 1990). Ultimately, the practical choice in adopting a methodology often occurs through a process of compromise between the owners of the evaluation process (ministries, governing agencies, funding councils) and those representing the interests of the affected academic community.

Australia's current national research assessment and the historic series of UK RAE exercises represent the two extreme poles. Australia's ERA (Excellence in Research for Australia ) adopts what can best be termed a "metrics-based expert review" model: here, the expert panels do not examine individual publications, and instead rely on bibliometrics in formulating their judgments. On the other hand the British RAE, since its inception in 1986, has adopted a peer review-based process. Still, the plan for the next UK exercise, scheduled for 2014 and named the Research Excellence Framework (REF), is to use informed peer review, where assessment outcomes are a product of expert review supported by bibliometric information.

One of the critical issues in any assessment is the choice of whether to evaluate the entire scientific portfolio of the institutions or a subset of the production. The REF, for example, will examine the work of selected research units and will not consider all their outputs – each of the selected units' members will submit a maximum of three or four of their highest quality works. The current Italian assessment, known as the VQR[2], is also based on a subset of products: universities are required to present, for each of their professors, the best three research works from the 2004-2010 period. On the other hand, in the Australian ERA, the institutions must submit the full research production of their entire research staff. The approach of evaluating all products clearly has advantages (Abramo and D'Angelo, 2011). However in systems involving peer review, for obvious reasons of costs and time it is clearly unthinkable to evaluate the entire output of a national research system. This fact results in undeniable penalties. First, it prevents providing any measure of productivity, the essential indicator of efficiency for any production system. Second, rankings are then sensitive to the size of whatever subset of output is evaluated, which seriously jeopardizes robustness of the underlying peer-review methodology. Third, it implies the necessity of selecting a subset of products (or researchers), which is not necessarily an efficient process, and which can well introduce elements of distortion that compromise the validity of the peer-review method. Fourth, it limits the functionality of the method, since it cannot be applied to all single researchers or research groups. As a consequence, universities do not receive performance rankings of their research staff, to in turn inform their internal selective funding. Finally, with respect to the bibliometric approach, peer-review implies very high costs and long times for execution, which limits its potential frequency of execution.

In this work we specifically wish to analyze the potential inefficiencies in the institutional process of selecting "best products" to submit for evaluation. The process appears to present significant risks that could nullify achievement of the very objectives of the evaluation. In fact, for institutional scores and rankings that descend from the submitted products to faithfully represent research quality, it is fundamental that the works submitted for evaluation truly be the institutions' best ones. If this were not the case, the evaluation exercises would produce rankings of quality that were distorted by an initial ineffective selection, with consequent inefficient allocation of resources. When the assessment exercises are based on a subset of the best products, as in the UK REF and the Italian VQR, it is thus fundamental that these truly represent the most valid part of the institutions' scientific portfolios, otherwise there is a risk of rewarding universities not for their true excellence, but for their ability or luck in selecting best products. In a preceding examination of the first Italian evaluation exercise (the 2001-2003 VTR)[3], Abramo et al. (2009) reported that the institutions demonstrated limited efficiency in selecting their products. The inexperience of the institutions, due to the fact that this was the first ever national evaluation, could have contributed to their difficulties in choosing best products. They would also encounter the objective difficulty of comparing and then choosing research outputs from different fields. In addition, selections could have been influenced by social factors, including the varying negotiating power of persons or groups. In the current VQR evaluation exercise, the risk of ineffective selection seems lessened by two new aspects. First, a prevalently bibliometric approach has been adopted for the hard sciences, meaning that the criteria for evaluating the products could be publicized in advance, and the choice of submissions to evaluation could then be based on quantitative parameters and objective data. In addition, the indication of the products to be submitted was entrusted first to the individual researchers, with each one asked to submit a list of their top products in order of quality, from which the institution could then forward the final number of submissions requested. However in reality, non-specialist academics found that it was not particularly easy to apply bibliometric indicators for support in selecting their best products. In addition, the fact that researchers were called to indicate their best products then gave them the possibility of boycotting or acting against the interests of their home institutions, by selecting worthless products or not providing any list at all (every product shortfall results in penalties to the institutional scores).[4]

---

[2] For information see http://www.anvur.org/?q=schema-dm-vqr-definitivo, last accessed 30 April 2013. As of April 2013, national panels of experts are evaluating the research products submitted by all universities.

Some of the universities engaged the current authors to assist them in the process of selecting products for submission to the VQR, in order to maximize their overall score. Starting from the original priority lists submitted by the individual researchers, it was then possible for us to observe the difference between the maximum potential score obtained through efficient selection of the best products and the score that would be obtained based on the top choices indicated by the individual researchers: the lesser the difference, the more efficient was the process of the individual researchers. In this work we measure the degree of efficiency in the selection process for three Italian universities. For each of these, we reconstruct the datasets of the publications proposed by the researchers and also develop datasets including publications that were not proposed but were still eligible for submission to evaluation.

The identification of errors by the researchers in indicating a different set of priority products than that suggested from bibliometric data and the calculation of the resulting difference in the evaluation scores permits identification of the penalties experienced for the overall institutions and each of their disciplines. The analysis is conducted using a purely bibliometric approach, in keeping with that used by the VQR.

The study is organized as follows: the next section describes the Italian VQR in general, while section 3 provides specific information on the requirements for the hard sciences portion of the exercise, as specified by the national expert panels. Section 4 highlights the critical issues inherent in the such national exercise requirements and presents the methodology for the current work. Section 5 presents the three case studies, first identifying the errors in product selection on the part of the researchers. Section 6 presents further analyses, with specific quantification of the penalization that the universities could experience at the level of overall score and for their individual disciplines. The concluding section summarizes the results and discusses their implications.

---

[3] The VQR was preceded by the VTR (Triennial Research Evaluation), conducted for the period 2001-2003: for information see http://vtr2006.cineca.it/index_EN.html (last accessed 30 April 2013).
[4] This problem occurred especially in a number of research institutions where recent reorganization had provoked union-management conflict.

## 2. The 2004-2010 Italian research assessment exercise, VQR

Until 2009, core government funding for Italian universities was input oriented: funds were distributed in a manner that would equally satisfy the needs of each and all, in function of institutional size and activities. The core funding, known as Ordinary Finance Funds (FFO) accounted for 56% of total university income. It was only following the first national evaluation exercise (2004-2006 VTR), that a minimal share, equivalent to 3.9% of total income, was assigned in function of assessments of research and teaching quality. The launch of the second Italian evaluation exercise (VQR) was preceded by vigorous debate, fed by heavy and indiscriminate cuts in financing to research and higher education that had been enacted by preceding governments. This debate saw opinions ranging from demands for more courageous action from policy makers in planning and implementing a true performance-based research funding system, able to induce improved performance at the various levels, to contrasting insistence on complete renunciation of any such initiative, or at least its serious revision. The VQR thus began in a period of heightened tensions. The Ministry of Education and Research (MIUR) entrusted implementation of the national exercise to the newly formed Agency for the Evaluation of University and Research Systems (ANVUR), which opened the evaluation process at the end of 2011. The exercise is intended to evaluate research activity carried out over the 2004-2010 period as conducted by:
- state universities;
- legally-recognized non-state universities;
- research institutions under the responsibility of the MIUR[5].

The objects of evaluation are the institutions, their macro-disciplinary areas and departments, but not the individual researchers. The results bear on two areas of future action: overall institutional evaluations will influence distribution of the merit-based share of FFO, while the evaluation of the macro-areas and departments could be used by the universities for the internal distribution of the acquired resources.

The evaluation of the overall institutions is to be determined by the weighted sum of a number of indicators: 50% based on a score for the quality of the research products presented and 50% derived from a composite of six other indicators (10% each for capacity to attract resources, mobility of research staff, internationalization and PhD programs, and 5% each for independent financing of research and overall improvement). Our current study focuses on the issues of the quality of the research products selected and the methodology for their evaluation.

Unlike the first exercise (2001-2003 VTR), which was entirely based on peer review, the 2004-2010 VQR is an evaluation exercise of a mixed type, primarily based on bibliometric analysis for some disciplines and on peer review for others. ANVUR nominated 14 evaluation panels (GEVs)[6] of national and international experts, one for each university disciplinary area (UDA) of the national academic system. Table 1 shows, for each UDA[7], the number of panelists in the GEV and the number of research staff on faculty in Italian universities as of 31 December 2011.

---
[5] Other public and private organizations engaged in research could participate in the evaluation by request, subject to fees.

| UDA/GEV | Panelists | Italian academic research staff |
| --- | --- | --- |
| 1 - Mathematics and computer science | 28 | 3,236 |
| 2 - Physics | 18 | 2,288 |
| 3 - Chemistry | 23 | 2,942 |
| 4 - Earth sciences | 9 | 1,086 |
| 5 - Biology | 38 | 4,903 |
| 6 - Medicine | 79 | 10,097 |
| 7 - Agricultural and veterinary sciences | 24 | 3,052 |
| 8 - Civil engineering and architecture | 28 | 3,623 |
| 9 - Industrial and information engineering | 40 | 5,288 |
| 10 - Ancient history, philology, literature and art | 42 | 5,345 |
| 11 - History, philosophy, pedagogy and psychology | 38 | 4,903 |
| 12 - Law | 37 | 4,887 |
| 13 - Economics and statistics | 36 | 4,832 |
| 14 - Political and social sciences | 13 | 1,746 |
| Total | 453 | 58,228 |

*Table 1: Numbers of evaluation panel members and research staff per university disciplinary area (UDA) as of 31/12/2011*

The institutions subject to evaluation were to submit a specific number of products for each unit of their research staff, in function of the rank of personnel and their period of activity over the seven years considered. The demand per university faculty member was up to a maximum of three products, while for research institutions the maximum expectation was six products per personnel member. ANVUR defined the acceptable products as: a) journal articles; b) books, book chapters and conference proceedings; c) critical reviews, commentaries, book translations; d) patents; e) prototypes, project plans, software, databases, exhibitions, works of art, compositions, and thematic papers.

Since results produced in collaboration with researchers in the same institution could only be presented once, the researchers were asked to identify a larger set of products than the minimal demand, from which the administration could then complete the selection of the number required for the VQR evaluation. The products were then submitted to the appropriate GEVs based on the researcher's identification of the field for each product. The GEVs are to judge the merit of each product as one of four values:

A = Excellent (score 1), if the product places in the top 20% on "a scale of values shared by the international community";
B = Good (score 0.8), if the product places in the 60%-80% range;
C = Acceptable (score 0.5), if the product is in the 50%-60% range;
D = Limited (score 0), if the product is in the bottom 50%.
The institutions are also subject to potential penalties:
- in proven cases of plagiarism or fraud (score -2);

---

[6] Acronym for Evaluation Groups
[7] Lists of panel members are available at http://www.anvur.org/sites/anvur-miur/files/gev_elenco_0.pdf, last accessed on 30 April 2013.

- for submissions that cannot be evaluated because they are not included in the products recognized by the GEV, lack appropriate documentation, or were not produced in the 2004-2010 period (score -1);
- for failure to submit the requested number of products (-0.5 for each missing product).

## 3. Criteria for evaluating product quality

Considering the characteristics of their disciplines, each GEV restricted the types of products permissible from the broader list and further defined the criteria and methods of evaluation. GEVs 10 to 14 opted exclusively for peer review, while GEVs 1 to 9 chose a mixed approach based on:
- bibliometric analysis, for articles indexed in the two major international services (Thomson Reuters WoS and Elsevier Scopus);
- peer review, for all other products, or where requested by the subject institution for indexed articles that were the result of work in an emerging area, of interdisciplinary character, or highly specialized.

For products subject to bibliometric evaluation, the judgments (A, B, C, D) are determined by a combination of two different indicators:
- a first indicator, named $I_R$, linked to the impact factor of the publishing journal (for those indexed in the WoS) or to the SCImago Journal Rank (for those indexed in Scopus);
- a second, named $I_C$, linked to the number of citations received by the article as of 31 December 2011.

Given the world distribution of these indicators for each subject category and year, the products are assigned to two classes, for journals and citations:
- class 1, if the article/journal places in the first quintile (top 80%-100%) of the reference world distribution;
- class 2, if the article/journal places in the 60%-80% range;
- class 3, if the article/journal places in the 50%-60% range;
- class 4, if the article/journal places in below the median of the reference world distribution.

Each product is thus attributed to one of the 16 possible combinations of the four classes for each indicator. Each GEV defined its own algorithm for deriving the ultimate judgment scores (A,B,C,D) from these combinations. Some GEVs further differentiated their algorithms on the basis of age of the products, meaning their year of publication. Figures 1 and 2 present the example of the classification matrices for the Chemistry UDA.

| $I_R \rightarrow$<br>$I_C \downarrow$ | 1 | 2 | 3 | 4 |
|---|---|---|---|---|
| 1 | A | A | A | IR |
| 2 | B | B | B | IR |
| 3 | IR | C | C | C |
| 4 | IR | D | D | D |

*Figure 1: Classification matrix for 2004-2008 products in the Chemistry UDA; IR = "evaluated by Informed Peer Review"*

| $I_R \rightarrow$<br>$I_C \downarrow$ | 1 | 2 | 3 | 4 |
|---|---|---|---|---|
| 1 | A | IR | IR | IR |
| 2 | A | B | C | D |
| 3 | A | B | C | D |
| 4 | IR | IR | IR | D |

*Figure 2: Classification matrix for 2009-2010 products in the Chemistry UDA. IR = "evaluated by Informed Peer Review"*

As seen in the Figure 1 matrix, the Chemistry GEV decided that for mature products (articles published in the 2004-2008 period), greater weight would be given to citations in determining the fnal merit judgment. Conversely, for recent products (publications from 2009-2010), judgment is based primarily on the impact of the publishing journal. The notation "IR" indicates situations where the values of the two indicators are inconsistent, with high citations in contrast to low values for journal impact, or vice versa. In these cases, the GEV decided to submit the products to informed peer review, meaning that the reviewers provided a judgment "informed" by the values of the bibliometric indicators. The other GEVs defined algorithms for merit judgment that are similar to the Chemistry example. The notable variations are:

- the Mathematics and computer science, Physics and Agricultural and veterinary science GEVs used only one algorithm for products of all ages;
- the Biology and Medicine GEVs chose the WoS as the only reference to be used in bibliometric evaluation and collapsed certain subject WoS categories into larger groupings, meaning that the reference distributions for both $I_R$ and $I_C$ are given by the merging of all world publications in the grouped categories;
- for $I_C$, in the Biology, Medicine, Earth science and Agricultural and veterinary science GEVs, the calculations for "articles" and "reviews" are based on distinct citation distributions;
- for $I_R$, the Mathematics and computer science and Industrial and information engineering GEVs drew on a combination of indicators from various sources and published a document with the resulting classification (1-4) of each journal; in particular, the Mathematics and computer science GEV established very stringent criteria for inclusion of journals in "class 1";
- the Agricultural and veterinary sciences GEV provided notice that review articles published in a pre-established list of journals would automatically be routed to evaluation by peer review.

All GEVs agreed on including self-citations for calculation of $I_C$.

The various evaluation criteria adopted by the GEVS were published at the end of February 2012, while the deadline for submission of products was set as 30 April 2012[8]. The institutions thus had two months to organize internally and select their products, in the light of the published information on the criteria.

The limited time available for selection of best products, the complexity of measuring the relative quality of their various products under the GEV criteria, particularly for non-specialists, and/or for those without the necessary bibliometric databases, as well as the strong variability in criteria across the GEVs, made the selection process remarkably difficult and led to notable disturbance in the institutions.

## 4. Selection of products on the part of the institutions

As noted in the introduction, for the scores produced from a research assessment exercise to represent the real value of the best institutional products, it is fundamental that the products submitted truly represent the top of what the institutions have produced. The methods of conducting the Italian VQR, but also the UK REF, do not guarantee this, since they delegate the responsibility of defining what is best and what is worst to the institutions (which in turn pass responsibility to the scientists). The transparency of criteria published by the GEVs only partially mitigates the risks of inefficient selection: as previously noted, certain scientists could decide to boycott or counter the interests of their own institution, willingly selecting worthless products or declining to present any at all; many others, acting with the best intentions, could have difficulty in discriminating the relative value of their products on the basis of such complicated algorithms and bibliometric data. As a starting point, many institutions in Italy have in fact not purchased licenses for access to WoS registries, while others have use of some but not others. There is thus a problem of information disparity for the subjects evaluated, with some being more "lucky" than others in having access to the bibliometric data for working out the world benchmark distributions (itself not at all an easy task). In effect, some universities assigned internal working groups to extract the data for working out these distributions[9]. Others turned to commercial services for access to bibliometric databases, others to experts in bibliometric evaluation. Certainly some institutions, undervaluing the risk of error and the rewards at stake, inappropriately delegated the entire process to their individual researchers. And though the major part seem to have assigned internal committees to support the process of product selection, it appears these did not correctly account for the GEV evaluation criteria or adherence to the requisites. In the following subsections we describe the methodology for demonstrating this statement.

---

[8] The deadline was later extended to 31 May and finally to 15 June 2012.

**4.1 Methodology and dataset**

To quantify the errors of selection committed by institutions in their submissions to the VQR (UDAs 1 to 9 only) we conduct case studies of three universities: one each from south, central and northern Italy. Table 2 indicates their total research staff, their staff in UDAs 1 to 9, and basic data on the submission of products for the VQR.

| University | Alpha | Beta | Gamma |
|---|---|---|---|
| Research staff (at the time of the call) | 744 | 524 | 1,704 |
| Of which in UDA 1-9 | 381 (51.2%) | 443 (84.6%) | 1,130 (66.3%) |
| Of which subject to evaluation | 367 (96.1%) | 424 (95.7%) | 1,114 (98.6%) |
| Of which submitted at least one product | 360 (98.1%) | 417 (98.3%) | 1,080 (96.9%) |
| Of which at least one was indexed in WoS | 348 (96.7%) | 385 (92.3%) | 911 (84.4%) |
| Products demanded for UDAs 1-9 | 1,058 | 1,212 | 3,221 |
| Number of products proposed by the researchers * | 7,481 | 1,656 | 19,802 |
| Of which were indexed in the WoS* | 5,060 (67.6%) | 1,474 (89.0%) | 12,193 (61.6%) |

*Table 2: Data on personnel and product submissions for the three case study universities*
*\* Figures actually indicate authorships and not individual products (a product is counted more than once if authored by more than one researcher in the same university).*

In the national system, each researcher or professor is identified as belonging to a specific scientific disciplinary sector (SDS, 370 in all)[10], grouped in 14 university disciplinary areas (UDAs, see list, Table 1).

The bibliometric data for the analysis are extracted from the Italian Observatory of

---

[9] In several cases the results from the internal workings of these committees appeared on the Internet, in what the authors consider an improper manner.
[10] The complete list is accessible at http://cercauniversita.cineca.it/php5/settori/index.php, last accessed 30 April 2013.

Public Research (ORP), a database developed and maintained by the authors and derived under license from Thomson Reuters WoS. Beginning from the raw data of the WoS and applying a complex algorithm for reconciliation of the author's affiliation and disambiguation of their precise identity, each publication is attributed to the university scientist or scientists that produced it (D'Angelo et al., 2011).

Data on staff members of each university and their SDS classifications are extracted from the database on Italian university personnel, maintained by the Ministry for Universities and Research[11].

## 4.2 Methodology

The dataset for each of the three universities examined consists of two distinct lists of products:
- Products proposed by the researchers (set A);
- Publications indexed in the WoS but not proposed by the researchers, attributed to them by analysis of ORP data (set B).

Beginning from these two data sets we develop a further three:
- $C = A \cup B$; the total set of works produced by the university's researchers;
- $D \subset A$; the subset of the three products proposed as "maximum priority" by each researcher[12];
- $E \subset C$; the subset of the three best products identified from set C, for each researcher.

These sets then permit quantification of proxies of the errors of selection in the recommendation of products by the researchers, as follows:

    I.   First, products "over-valued" by the researcher: $D - (D \cap E)$. This type of error is committed when the researcher recommends the product as high priority (1, 2 or 3) when in reality it places below third place in their overall scientific portfolio, when ranked under the criteria for the relevant GEV.

    II.  Products "undervalued" by the researcher: $E \cap (A - D)$. This second type of error is committed when the researcher proposed lower priority for the product (4, 5, etc.) but it actually places in the first three positions in the ranking for their overall portfolio, under the criteria set by the GEV.

    III. Products "omitted" by the researcher: $E - (A \cap E)$. This third type of error occurs when a publication indexed in the WoS is not proposed by the researcher, while under the criteria for the GEV it would be in the top three places of the individual's 2004-2010 publications ranking.

We then apply the criteria defined by the GEVs, proceeding to bibliometric analysis only for those products listed in the WoS and adopting certain assumptions concerning those directed to peer review or "informed peer review", as detailed below.
- For products that were directed to informed peer review under the GEV procedures, we assign a score of 0.5. These are the products where the values of the two bibliometric indicators are inconsistent (high impact factor and low number of citations or vice versa). We hypothesize that the reviewers would adopt a judgement with "intermediate" modal value under the scale available.
- For products indexed in the WoS but without impact factor (usually conference proceedings) we assign values of: i) nil, if intended for Biology and Medicine (where the GEVs explicitly notified that they would be given nil score); ii) 0.5 if intended for Information engineering (since in this field conference procedings are considered to have value); iii) 0.25 in all other cases.
- For products not indexed in the WoS we assign a score of 0.25.

---

[11] http://cercauniversita.cineca.it/php5/docenti/cerca.php, last accessed 30 April 30 2013

[12] For simplicity in language we consistently indicate that the institutions were expected to submit "three" products for each researcher, although the precise number depended on the academic rank and the period that the faculty members were on staff over the years subject to evaluation (see full explanation, Section 2).

The assumption of 0.25 score for the final two points considers that in the hard sciences there are only a limited number of resarchers that produce less than three publications indexed in the WoS. The resort to submission of non-indexed publications indicates that the researcher is quite unproductive, and we also assume the quality of their works is relatively modest.

We further assume that potential products proposed by the researchers as 1, 2 or 3 priority, which are indexed in Scopus but not in WoS, would not be better than those contained in subset E. Finally, we assume that products not censused in either WoS or Scopus and not submitted by the researcher are not better than those contained in E.

All of the above assumptions should be recalled for proper interpretation of the results of the analysis, as reported in the following sections.

## 5. Errors in selection made by the three universities under analysis

Table 3 presents, for each of the three universities: i) the number of WoS-indexed publications proposed by the researchers: ii) the number of WoS publications that were not proposed; iii) the ratio between the "not-proposed" publications and the total WoS-indexed publications for the university.

| | Alpha | | Beta | | Gamma | |
|---|---|---|---|---|---|---|
| UDA | P | NP | P | NP | P | NP |
| 1 - Mathematics and computer science | 226 | 274 (55%) | 45 | 29 (39%) | 435 | 115 (21%) |
| 2 - Physics | 82 | 28 (25%) | 77 | 209 (73%) | 1,298 | 170 (12%) |
| 3 - Chemistry | 288 | 77 (21%) | 46 | 86 (65%) | 1,831 | 90 (5%) |
| 4 - Earth sciences | - | - | 16 | 25 (61%) | 272 | 7 (3%) |
| 5 - Biology | 874 | 437 (33%) | 287 | 888 (76%) | 1,528 | 99 (6%) |
| 6 - Medicine | 3,356 | 1,679 (33%) | 479 | 1,953 (80%) | 4,589 | 588 (11%) |
| 7- Agricultural and veterinary sciences | 112 | 99 (47%) | 121 | 218 (64%) | 517 | 60 (10%) |
| 8- Civil engineering and architecture | - | - | 95 | 122 (56%) | 290 | 82 (22%) |
| 9 - Industrial and inform. engineering | 122 | 131 (52%) | 308 | 901 (75%) | 1,433 | 775 (35%) |
| Total | 5,060 | 2,725 (35%) | 1,474 | 4,431 (75%) | 12,193 | 1,986 (14%) |

*Table 3: Publications\* indexed in the WoS that were proposed (P) and not proposed (NP) for VQR submission, for the three universtities examined (in parentheses the ratio of NP to P + NP)*

\* *Figures actually indicate authorships and not individual products (a product is counted more than once if authored by more than one researcher of the same university).*

In the last line, the data indicate that at university Alpha the researchers omitted over a third of their WoS-indexed scientific portfolio (35%) from the set they proposed. For university Gamma the omissions drop to 14% while soaring to 75% for university Beta. The reasons for this disparity can be traced to the different procedures adopted to respond to the VQR demands. While universities Alpha and Gamma asked each researcher to present the list of their entire 2004-2010 production, Beta requested only their best products. Lines 8 and 9 in the table show that researchers at Alpha and Gamma thus presented 6 to 7 times more publications than what was ultimately necessary for submission to the GEVs, while at Beta the publications proposed by the researchers (1,656) are little more than was ultimately due (1,212).

We now proceed to identify and quantify the errors and omissions of each university through analysis and comparison of sets D and E (Table 4). Each university selected products that were unacceptable under the GEV criteria and were thus penalized in their scores. Alpha, with 2.7% of products unacceptable, was the least attentive to this aspect. There were also substantial numbers of publications submitted that would receive nil scores under the bibliometric criteria set by the GEVs: 85 for Alpha (8% of the products in set D), 100 for Beta (9%) and 456 for Gamma (15%). These products could have been substituted with others, where available, meeting the necessary criteria for evaluation by peer review: in such case, the probability of receiving a score greater than 0 would be non-nil. Another type of error is the overestimation of the value of certain products compared to others in the lists provided by the researchers. This occurs for a third of Alpha's products, almost a quarter of Beta's and 30% of the products at Gamma. Integrating the lists provided by the researchers with the entire portfolio of their WoS publications and for each individual extracting the best products by GEV criteria (set E), we immediately note that the number of products with nil bibliometric score is much less compared to set D: nil scores are now 20 compared to the original 85 for Alpha, 33 compared to 100 for Beta and 43 to 456 for Gamma. For Alpha, 35.6% of the publications present in set E are not included in D: this overall figure represents both a "missing" share from researchers that proposed their WoS products but under-valued them in their recommendations (25%) and another relative to WoS publications that researchers completely omitted from their recommendations (10.6%).

The data for Gamma are essentially the same as for Alpha, however as expected given the procedures adopted, the situation for Beta is different: here the incidence of set E products not present in D is 29%, but the composition is much different than for Alpha and Gamma, with 5.8% originating from products under-evaluated and 23.2% from products completely omitted.

|  | Alpha | Beta | Gamma |
|---|---|---|---|
| Products to be submitted | 1,058 | 1,212 | 3,221 |
| Number of products in set D | 1,030 | 1,101 | 2,945 |
| Of which inadmissible | 27 | 4 | 12 |
| Of which with nil bibliometric score | 85 | 100 | 456 |
| Of which over-valued | 347 (33.7% of D) | 265 (24.1% of D) | 890 (30.2% of D) |
| Number of products, set E | 1,028 | 1,136 | 2,993 |
| Of which with nil bibliometric score | 20 | 33 | 43 |
| Of which undervalued | 257 (25.0% of E) | 66 (5.8% of E) | 856 (28.6% of E) |
| Of which omitted from submission | 109 (10.6% of E) | 264 (23.2% of E) | 149 (5.0% of E) |

*Table 4: Errors and omissions in selection of "best products" by researchers for the three universities examined*

## 6. Impact of errors in selection on national scores

In the preceding section we observed that the recommendations from researchers intended to guide the selection of products for their institutional submissions to evaluation are characterized by significant rates of error. As shown in Table 5, the average scores of the recommended products (set D) are in fact markedly lower than the average scores for set E. For Alpha, the average difference in score is 0.21 points, representing a 34% increase from set D to set E. If we exclude the products directed to peer review, limiting the analysis to those that can be given a definite score based on the GEV-defined criteria, the difference between D and E remains at 0.18 points (+24%). For university Gamma, calculated in similar manner, the increases in average value are respectively 31% and 41%. Beta shows lower differences in average score, respectively 17% and 15%.

|  | Alpha | | Beta | | Gamma | |
|---|---|---|---|---|---|---|
|  | All products | Those with definite score | All products | Those with definite score | All products | Those with definite score |
| Products in set D | 0.61 | 0.75 | 0.66 | 0.78 | 0.52 | 0.64 |
| Products in set E | 0.83 | 0.93 | 0.77 | 0.91 | 0.67 | 0.91 |
| Difference | 0.21 (+34%) | 0.18 (+24%) | 0.11 (+17%) | 0.12 (+15%) | 0.16 (+31%) | 0.26 (+41%) |

*Table 5: Average score of products proposed for evaluation by the three universities examined*

We now ask what will be the consequences of these errors, recalling that at the close of the evaluation exercise, the reward-based financing provided by the national government depends on the overall score of the products from each university.

To respond to the question we simulate the procedures for the VQR evaluation of research quality. We also do this inserting a further condition of the exercise: that products with co-authors from the same institutions could only be presented once (all "multiple" submissions were considered completely void). This rule was a further cause of error, since refinement of product selection in keeping with the requirement could only be accomplished by central coordination at the institutional level. We simulate the process of final selection through three hypothetical scenarios in increasing levels of complexity:

- In Scenario 1, the hypothesis is that the universities would select the products for submission from set A, on the basis of the priorities indicated by the researchers. In cases of co-authorship, the contested product is assigned to the researcher that assigns the higher priority.
- In Scenario 2, the universities would again select the products from set A but not according to the priorities suggested by the researchers, and instead on the basis of bibliometric evaluation referring to the GEV criteria. In this case co-authorship is resolved in favor of the researcher with remaining products[13] that would achieve a lower score.
- In Scenario 3, the selection process is similar to 2, but begins from the total set of products (C) obtained by combining the set proposed by the researchers (A) with WoS-indexed publications that were not proposed but were authored by the same individuals (B).[14]

Table 6 presents the results from the simulations for university Alpha, with the total scores under the different scenarios seen in the last line: we observe that the difference between scenarios 1 and 2 is +25.9%, while the difference between scenarios 1 and 3 reaches +32.2%. The first figure shows that the university would be particularly penalized if the process of product selection was completely entrusted to the judgment of the researchers, instead of determining the bibliometric score of each of the recommended products under GEV-defined criteria. A further significant increase would be obtained by taking the recommendations of the researchers, adding the WoS publications for the 2004-2010 period that they did not recommend, and then choosing the best under GEV criteria. Medicine would be the area most affected: for this UDA the difference between scenarios 1 and 3 is +38.8%. The least improved UDA is Physics, with optimization of the selection process obtaining a 13.5% improvement between scenarios 1 and 2, and no further benefit from integrating the researchers' remaining WoS publications.

---
[13] This refers to products authored by the researchers but not selected for submission.
[14] In reality no Italian institution independently prepares exhaustive databases of all the products of its researchers, or of their publications indexed in the WoS. The simulation of scenarios 2 and 3 is possible using the ORP databases.

| UDA | Products to submit | Scen. 1 | Scen. 2 | Scen. 3 | 1 vs 2 | 2 vs 3 | 1 vs 3 |
|---|---|---|---|---|---|---|---|
| 1 - Mathematics and computer science | 110 | 55.6 | 67.9 | 69.6 | 22.0% | 2.6% | 25.2% |
| 2 - Physics | 18 | 13.0 | 14.8 | 14.8 | 13.5% | 0.0% | 13.5% |
| 3 - Chemistry | 24 | 19.4 | 22.4 | 22.2 | 15.5% | -0.9% | 14.4% |
| 5 - Biology | 184 | 111.2 | 133.5 | 140.6 | 20.1% | 5.4% | 26.5% |
| 6 - Medicine | 618 | 338.7 | 443.5 | 470.3 | 30.9% | 6.0% | 38.8% |
| 7- Agricultural and veterinary sciences | 62 | 37.8 | 42.7 | 44.7 | 13.0% | 4.6% | 18.1% |
| 8- Civil engineering and architecture | 3 | 0.8 | 0.8 | 0.8 | 0.0% | 0.0% | 0.0% |
| 9 - Industrial and inform. engineering | 39 | 22.5 | 28.6 | 28.6 | 27.2% | 0.2% | 27.4% |
| Total | 1,058 | 598.9 | 753.9 | 791.4 | 25.9% | 5.0% | 32.2% |

*Table 6: Total scores deriving from simulations of the national evaluation exercise, under three different scenarios (university Alpha)*

Table 7 presents the three results of the simulations for university Beta. As previously noted, this university requested its staff to present just few more publications than ultimately demanded[15]. Thus in the comparison between scenarios 1 and 2, having less margin for improvement, we observe a difference of 6.8% between the total scores - much lower than the case for Alpha. Still, in the comparison between scenarios 2 and 3 we observe a notable increase in total score (+15.3%), deriving from the integration of the starting dataset with the WoS products that the researchers omitted from their recommendations. Between the two extreme scenarios (1 and 3), there is certainly an evident difference (+23.1%), with particularly critical cases in the two smaller UDAs: Chemistry (+83.1%) and Earth sciences (+54.5).

For university Gamma, the results of the simulation are similar to those for Alpha (Table 8). The difference in overall score between scenarios 1 and 3 is +28.5%. The major contribution to this figure is from error committed by researchers in indicating the priority of the products they proposed for selection: (+21.8% between scenarios 1 and 2). The error due to complete omission of products indexed in the WoS from the recommendations is more contained (+5.5% between scenarios 2 and 3). Again here, as for Alpha, the error is not distributed equally between the UDAs. The maximum value of error is again observed in the largest UDA, Medicine, where the delta score between the two extreme scenarios is 38.8%. In the other large UDAs (Biology, Civil engineering and architecture, Industrial and information engineering, all of which demanded over 500 submissions), the difference between the scenario 1 and 3 scores is systematically over 30%.

---

[15] The university actually requested the "extra" publications only to permit resolution of potential problems of co-authorship.

| UDA | Products to submit | Scen. 1 | Scen. 2 | Scen. 3 | 1 vs 2 | 2 vs 3 | 1 vs 3 |
|---|---|---|---|---|---|---|---|
| 1 - Mathematics and computer science | 41 | 18.5 | 19.0 | 22.3 | 3.0% | 17.4% | 20.9% |
| 2 - Physics | 47 | 30.1 | 30.3 | 36.0 | 0.8% | 18.6% | 19.6% |
| 3 - Chemistry | 39 | 13.1 | 14.8 | 23.9 | 13.4% | 61.5% | 83.1% |
| 4 - Earth sciences | 16 | 5.5 | 7.6 | 8.5 | 38.2% | 11.8% | 54.5% |
| 5 - Biology | 208 | 135.3 | 144.2 | 162.7 | 6.5% | 12.9% | 20.3% |
| 6 - Medicine | 356 | 216.3 | 230.0 | 260.7 | 6.4% | 13.3% | 20.6% |
| 7- Agricultural and veterinary sciences | 117 | 54.5 | 57.4 | 62.6 | 5.3% | 9.1% | 14.9% |
| 8- Civil engineering and architecture | 138 | 32.8 | 33.0 | 45.3 | 0.6% | 37.1% | 38.0% |
| 9 - Industrial and inform. engineering | 250 | 129.4 | 142.2 | 160.1 | 9.9% | 12.6% | 23.7% |
| Total | 1,212 | 635.3 | 678.4 | 782.0 | 6.8% | 15.3% | 23.1% |

*Table 7: Total scores core deriving from simulations of the national evaluation exercise, under three different scenarios (university Beta)*

| UDA | Products to submit | Scen. 1 | Scen. 2 | Scen. 3 | 1 vs 2 | 2 vs 3 | 1 vs 3 |
|---|---|---|---|---|---|---|---|
| 1 - Mathematics and computer science | 198 | 76.3 | 87.7 | 94.9 | 15.0% | 8.3% | 24.5% |
| 2 - Physics | 170 | 114.0 | 128.9 | 128.2 | 13.1% | -0.6% | 12.4% |
| 3 - Chemistry | 250 | 165.4 | 189.0 | 192.5 | 14.3% | 1.8% | 16.4% |
| 4 - Earth sciences | 93 | 50.9 | 55.1 | 55.9 | 8.3% | 1.5% | 9.8% |
| 5 - Biology | 475 | 191.9 | 245.1 | 255.0 | 27.8% | 4.0% | 32.9% |
| 6 - Medicine | 856 | 362.7 | 480.7 | 503.3 | 32.5% | 4.7% | 38.8% |
| 7 - Agricultural and veterinary sciences | 292 | 110.7 | 131.6 | 135.0 | 18.9% | 2.6% | 22.0% |
| 8 - Civil engineering and architecture | 444 | 52.9 | 63.4 | 71.2 | 20.0% | 12.2% | 34.6% |
| 9 - Industrial and inform. engineering | 443 | 199.7 | 232.0 | 266.3 | 16.2% | 14.8% | 33.4% |
| Total | 3,221 | 1,324.2 | 1,613.35 | 1,702.15 | 21.8% | 5.5% | 28.5% |

*Table 8: Total scores core deriving from simulations of the national evaluation exercise, under three different scenarios (university Gamma)*

## 7. Conclusions

For the Italian university system, the current national evaluation exercise (2004-2010 VQR) is an event of vital importance, since the results will support the process of allocating a "reward" share of core government funding.

One of the critical methodological issues in the VQR, and in national assessment exercises in general, concerns the choice of whether to evaluate the entire scientific portfolio of the institutions or a subset composed of the best products. In the VQR, for each member of their staff, the institutions were asked to submit a certain number of products for evaluation. The risks both for the funding agency and the individual institutions are then that the difficulty in carrying out efficient selections could lead to final scores and relative assignments of selective funding that do not represent the true quality of research conducted in the institutions. For this, in the current work we have examined three case studies of Italian universities. Through a simulation approach that closely corresponds to the VQR procedures, we evaluate the rate of error in the selection of products by the three universities, and the penalization that they experience in terms of overall score and by individual discipline.

We observed that in a small but notable number of cases the researchers recommended submission of products that did not meet the criteria for consideration under the VQR. In a much greater number of cases, the researchers tended to over-evaluate some of the works in their lists of recommended scientific products, placing lesser products among those they rated as best for submission. In numerous further cases the researchers recommended lists of products that completely omitted other works, by the same individual, which were superior to the ones proposed. The details of design of national exercises can render the selection of products still more complex: in the case of the Italian VQR, for example, further difficulties (and error) arose from the rule that products coauthored by researchers from the same institution could not be submitted twice. After observing the errors in the actual recommendations of works by the researchers, we then simulated three scenarios corresponding to successively more sophisticated optimization of the university product selection procedures.

In terms of the impact of errors in selection for the final scores of quality, the results of the simulation demonstrate that the universities are particularly penalized in the selection of their products if they adhere exclusively to the proposals from the researchers, instead of choosing from among them on the basis of the criteria defined by the VQR panels. A further significant increase in score can be obtained by the universities if they choose from all the scientific production of the researchers, rather than only the lists of works that the researchers recommend to their home institution.

Given our analysis, we conclude that it is highly probable that the national ranking lists produced by the exercise will be strongly distorted by the types of errors noted above, with negative consequences for the subsequent distribution of financing, and further effects completely contrary to the logic of incentivizing for which the VQR was conceived.

Some universities will certainly be better equipped to deal with such requirements in calls for submissions – indeed, the three case studies indicate clearly different rates of error, depending on the process of selection implemented by the universities. The errors in selection on the part of the universities further undermine the already weak methodological structure at the base of the VQR, as illustrated in the introduction. The only hope is that the distortions introduced in the final rankings by all of these factors will somehow cancel out, perhaps by chance. Only in this way will it be possible to limit the damage caused by allocation of resources based on distorted rankings.

However the rebirth of academic systems such as Italy's depends precisely on implementing incisive and effective reward measures, such as might have been hoped for from the VQR. The authors thus question whether it would have been opportune, at least for the disciplines where bibliometrics are appropriate, to evaluate the entire scientific portfolio of the university. Such a strategy would involve less expenditure of funds and time, would avoid errors due to selection, and obtain real productivity measures on which to base not only financial incentives for the universities, but also internal incentivizing mechanisms, distributing "revenue to revenue generators".

**References**

Abramo, G., D'Angelo, C.A. (2010). Evaluating research: from informed peer review to bibliometrics, *Scientometrics*, 87(3), 499-514.

Abramo, G., D'Angelo, C. A., Caprasecca, A. (2009). Allocative efficiency in public research funding: Can bibliometrics help? *Research Policy*, 38(1), 206–215.

Abramo, G., D'Angelo, C. A., Viel, F. (2010). Peer review research assessment: a sensitivity analysis of performance rankings to the share of research product evaluated, *Scientometrics*, 85(3), 705–720.

Butler, L., McAllister, I. (2007). Metrics or peer review? Evaluating the 2001 UK Research assessment exercise in political science, *Political Studies Review*, 7(1), 3–17.

Council of Canadian Academies (2012). *Expert Panel on Science Performance and Research Funding Informing research choices: indicators and judgment /The Expert Panel on Science Performance and Research Funding*. ISBN 978-1-926558-42-4.

Department of Education. Science and Training (DEST) 2005. Research Quality Framework: Assessing the Quality and Impact of Research in Australia. *Advanced Approaches Paper*, Government of Australia, Canberra.

ERA (Excellence in Research for Australia). http://www.arc.gov.au/era/default.htm. Last accessed on July 15, 2013.

Franceschet, M., Costantini, A. (2009. The first Italian Research Assessment Exercise: a bibliometric perspective. *Journal of Informetrics*, 5(2), 275–291.

Geuna, A., Martin, B.R. (2003). University research evaluation and funding: an international comparison. *Minerva*, 41(4), 277–304.

Herbst, M. (2007). Financing Public Universities. *Higher Education Dynamics*, 18.

Hicks, D. (2012). Performance-based university research funding systems. *Research Policy*, 41(2), 251–261.

Horrobin, D.F. (1990). The philosophical basis of peer review and the suppression of innovation. *Journal of the American Medical Association*, 263(10), 1438–1441.

Jordan, G.B., Malone, E.L. (2001). Performance Assessment. In US Department of Energy Office of Science and Office of Planning and Analysis (Ed.). *Management Benchmark Study*. Washington (DC), US Department of Energy.

Moxham, H., Anderson, J. (1992). Peer review. A view from the inside. *Science and Technology Policy*, 5(1), 7–15.

OECD (2010). Performance-based Funding for Public Research in Tertiary Education Institutions: Workshop Proceedings. OECD Publishing. http://dx.doi.org/10.1787/9789264094611-en. Last accessed on July 15, 2013

REF (Research Excellence Framework) 2009. http://www.hefce.ac.uk/pubs/hefce/2009/09_38/#exec. Last accessed on July 15, 2013.

Research Excellence Framework (REF) 2014. http://www.ref.ac.uk/ Last accessed on July 15, 2013

Schubert, T. (2009). Empirical observations on New Public Management to increase efficiency in public research—Boon or bane? *Research Policy*, 38(8), 1225–1234.

Stoker, G. (2006). Public value management: a new narrative for networked governance? *The American Review of Public Administration*, 36(1), 41–57.

Van der Most, F. (2010) Use and non-use of research evaluation: A literature review. In *CIRCLE Electronic Working Paper Series. WP2010/16* (pre-print version of a paper submitted to *Research Evaluation*).

VQR (2012). Valutazione della Qualità della Ricerca (2004-2010). http://www.anvur.org/sites/anvur-miur/files/bando_vqr_def_07_11.pdf. Last accessed on July 15, 2013